\documentclass[
aps,%
12pt,%
final,%
notitlepage,%
showkeys,%
oneside,%
onecolumn,%
nobibnotes,%
nofootinbib,% 
superscriptaddress,%
noshowpacs,%
centertags]%
{revtex4}

\newcommand{\kms}{km\,s$^{-1}$}

\input maik.sty 
\usepackage{xcolor}
\usepackage[utf8x]{inputenc}
\def\refitem#1{\relax}

\begin{document}

\selectlanguage{english}

\title{\MakeTextUppercase{H$^{13}$CN-HN$^{13}$C intensity ratio as a temperature indicator of interstellar clouds}}

\author{\firstname{A.~G.}~\surname{Pazukhin}}
\email{pazukhinandrey@bk.ru}
\author{\firstname{I.~I.}~\surname{Zinchenko}}
\email{zin@ipfran.ru}
\affiliation{Federal Research Center Institute of Applied Physics RAS, Nizhny Novgorod, Russia}
\affiliation{Lobachevsky State University of Nizhny Novgorod, Nizhny Novgorod, Russia}

\author{\firstname{E.~A.}~\surname{Trofimova}}
\email{tea@ipfran.ru}
\affiliation{Federal Research Center Institute of Applied Physics RAS, Nizhny Novgorod, Russia}

\author{\firstname{C.}~\surname{Henkel}}
\email{chenkel@mpifr-bonn.mpg.de}
\affiliation{Max Planck Institute for Radio Astronomy, Bonn, Germany}
\affiliation{King Abdulaziz University, Jeddah, Saudi Arabia}

\begin{abstract}

With the 30-m IRAM radio telescope, we observed several massive star forming regions at wavelengths of 3-4 and 2 mm. The temperature of the gas in the sources was estimated from the lines of CH$_{3}$CCH and from the transitions of the NH$_3$ molecule obtained during observations at the 100-m radio telescope in Effelsberg. As a result, {a correlation between the integrated intensity ratios of the $J=1-0$ transitions of H$^{13}$CN and HN$^{13}$C and the kinetic temperature has been obtained.} The obtained results allow us to propose the use of the intensity ratio H$^{13}$CN-HN$^{13}$C as a possible temperature indicator of interstellar clouds. We {also} compared the obtained estimates of the kinetic temperature with the dust temperature $T_{dust}$. As a result, no significant correlation was found. 

\end{abstract}

\received{September 02, 2022}
\revised{October 10, 2022}
\accepted{October 20, 2022}


\maketitle

\section{Introduction}

The hydrogen cyanide molecule HCN and {its} isomer HNC are widely distributed in the interstellar medium.
It is known that the HCN/HNC abundance ratio strongly depends on the kinetic temperature, for example, it was found in~\cite{Jin15} that the abundance ratio in high mass protostellar objects is 4, in hot ultracompact H~II regions the average value is 9.
In~\cite{Hacar20} {it was} proposed to use the intensity ratio of HCN to HNC line as a temperature indicator based on observations of the integral shaped filament in Orion.

The main pathway for the formation of HCN and HNC isomers is the dissociative recombination of the HCNH$^+$ ion with an electron:
\begin{eqnarray}
    {\rm HCNH}^{+} \ + \ {\rm e}^{-} \ \to
    \begin{cases}
    \ {\rm HCN} \ + \ {\rm H} \, \\
    \ {\rm HNC} \ + \ {\rm H} \,
    \end{cases}\label{rx:hcnh+}.
\end{eqnarray}
This reaction has an approximately equal branching ratio~\cite{Mendes}, and the abundance differences between HCN and HNC are largely determined by the destruction and isomerization reactions of HNC. These include the following reactions (e.g.~\cite{Graninger}):
\begin{eqnarray}
    {\rm HNC} \ +\ {\rm H} \ \to\ {\rm HCN} \ +\ {\rm H} \label{rx:hnc/h}\\
    {\rm HNC} \ +\ {\rm O} \ \to\ {\rm CO} \ +\ {\rm NH} \label{rx:hnc/o}.
\end{eqnarray}
The energy barrier for the reaction~(\ref{rx:hnc/h}) is 200~K~\cite{Graninger}, for the reaction~(\ref{rx:hnc/o}) it is 20~K, which determines the dominant role of  reaction~(\ref{rx:hnc/o}) at low temperatures of the order of 50~K~\cite{Hacar20}. However, the classical calculated energy barriers are 1200~K and 2000~K, respectively (see details in~\cite{Graninger}).

\section{Observations and data reduction}

\subsection{Observations at the 30-m radio telescope of the Institute of Millimeter Radio Astronomy (IRAM)}

\begin{table}
\setcaptionmargin{0mm}
\onelinecaptionstrue
\captionstyle{flushleft}
\caption{List of sources } \label{tab1}
\bigskip
\begin{tabular}{l|c|c|c|c|c}
            \hline
            Object & $\alpha$ (2000) & $\delta$ (2000) & $V_{lsr}$ & $d$ & Note \\
             & &  & \kms & kpc & \\
            \hline
            L~1287 & 00$^h$36$^m$47.5$^s$ & 63$^o$29$^{\prime}$02.1$^{\prime\prime}$ & -17.7& 0.93 & G121.30+0.66, IRAS~00338+6312\\
            S~187 & 01~23~15.4 & 61~49~43.1 & -14.0& 1.0& G126.68–0.81, IRAS~01202+6133 \\
            S~231 & 05~39~12.9 &35~45~54.0 &-16.6 & 2.3 & G173.48+2.45, IRAS~05358+3543 \\
            DR~21(OH) & 20~39~00.6 & 42~22~48.9 & -03.8 & 1.5 & G81.72+0.57 \\
            NGC~7538 & 23~13~44.7 & 61~28~09.7 &-57.6 & 2.8 & G111.54+0.78, IRAS~23116+6111 \\
            \hline
           
        \end{tabular}
\begin{flushleft}
    \scriptsize \textbf{Note.} Distances to sources are quoted from ~\cite{Rygl10, Fich84, Rygl12}
\end{flushleft}
\end{table}

\begin{table}
\setcaptionmargin{0mm}
\onelinecaptionstrue
\captionstyle{flushleft}
\caption{Observed molecular lines} \label{tab2}
\bigskip
\begin{tabular}{l|c|r|r}
        \hline
        Molecule & Transition & Frequency, MHz& $E_{u}/k$, K \\
        \hline
            NH$_3$ & $(1,1)$ & 23694.495 & 23.4 \\
              & $(2,2)$ & 23722.634 & 64.9 \\
            CH$_{3}$CCH & $5_{3}-4_{3}$ & 85442.601 & 77.3 \\
        & $5_{2}-4_{2}$ & 85450.766 & 41.2 \\
        & $5_{1}-4_{1}$ & 85455.667 & 19.5 \\
        & $5_{0}-4_{0}$ & 85457.300 & 12.3 \\
        H$^{13}$CN & $1-0$ & 86339.921 & 4.1 \\
        HN$^{13}$C & $1-0$ & 87090.825 & 4.2 \\
        HCN & $1-0$ & 88631.602 & 4.3 \\
        HNC & $1-0$ & 90663.568 & 4.4 \\
        CH$_{3}$CCH & $9_{3}-8_{3}$ & 153790.772 & 101.9 \\
        & $9_{2}-8_{2}$ & 153805.461 & 65.8 \\
        & $9_{1}-8_{1}$ & 153814.276 & 44.1 \\
        & $9_{0}-8_{0}$ & 153817.215 & 36.9 \\
        \hline
    \end{tabular}
\end{table}

In September 2019, with the 30-m radio telescope of the Institute of Millimeter Radio Astronomy (IRAM), we observed several massive star forming regions at wavelengths of 2 and 3--4~mm (as part of project 041-19). The list of sources is given in Table~\ref{tab1}. In this paper, a part of the obtained data is discussed. Table~\ref{tab2} contains the corresponding list of molecular lines. The transition frequency and upper level energy are taken from the CDMS\footnote{\url{http://cdms.de}} catalogue.

The full beam width at half maximum at the discussed frequencies ranged from $\sim30^{\prime\prime}$ to $\sim16^{\prime\prime}$. The antenna temperature $T^*_A$ was reduced to the main beam temperature $T_{mb}$, using the main beam efficiency $B_{eff}$, which was determined by the Ruze's equation in accordance with the IRAM recommendations and ranged from 0.72 to 0.82. The minimum system noise temperature was $\sim 100$~K in the 3~mm range and $\sim 200$~K in the 2~mm range.

Observations were carried out by the method of continuous mapping (OTF, On-The-Fly) of a $200^{\prime\prime}\times 200^{\prime\prime}$ area in full power mode. The reference position was chosen with a shift of 10$^{\prime}$ in right ascension. In some extended sources, {i.e. DR~21(OH) and NGC~7538,} two partially overlapping areas were observed. The pointing accuracy was checked periodically by observations of nearby continuum sources.

\subsection{Observations at the {Max-Planck-Institute} for Radio Astronomy {with the} Effelsberg 100-m radio telescope} 
{On 9 December 2019 we observed with the 100-m telescope near Effelsberg (Germany) the H$_2$O maser} transition at a frequency of 22~GHz, as well as the ammonia inversion lines {$(J,K)=\,$}(1,1), (2,2) and (3,3). The full beam width at half maximum was $\sim40^{\prime\prime}$.
The measurements were carried out by the method of continuous mapping using a $K$-band receiver in a secondary focus with a dual bandwidth of 300~MHz, including the aforementioned H$_2$O lines in one band and NH$_3$ in the other band. $5^\prime \times 5^\prime$ maps were obtained at a scanning rate of 20$^{\prime\prime}$ per second in right ascension; intervals between scans were 15$^{\prime\prime}$. The reference position was shifted by +15$^\prime$ in azimuth. Weather conditions included light rain with low wind speeds ($\sim$2~m\,s$^{-1}$).

The source NGC~7027 was used for calibration with a flux density of 5.5~Jy at 22~GHz~\cite{Ott94}. The antenna temperature $T^*_A$ was obtained by multiplying the observed intensities by $T_{cal}$ and taking into account atmospheric absorption\footnote{\url{http://eff100mwiki.mpifr-bonn.mpg.de}}.

\subsection{Data reduction}
The GILDAS/CLASS software\footnote{\url{http://www.iram.fr/IRAMFR/GILDAS}} was used for data reduction. 
The {IRAM-30m} and {Effelsberg-100m} datasets were reduced to the same spatial resolution of 40$^{\prime\prime}$. After baseline subtraction and smoothing, the spectral resolution for the {Effelsberg-100m} data was $\sim$0.46~\kms.

In the analysis, integrated intensity maps ($I = \int T_{mb}dV$ in units~K$\cdot$\kms) were used in the velocity range [$V_{lsr}$-10, $V_{lsr} $+10] for HCN, H$^{13}$CN and [$V_{lsr}$-4, $V_{lsr}$+4] for HNC, HN$^{13}$C. It should be noted that two velocity components $\sim -4$ and $\sim 0$~\kms\ are observed in the source DR~21(OH) (see details in~\cite{Schneider}). In the reduction, the components were separated, {and only the $\sim -4$~\kms\ component has been used for the analysis, since it is stronger and is detected throughout the source.}

\section{Results}
\subsection{Kinetic temperature from observations of CH$_{3}$CCH}

{In~\cite{Askne84, Bergin} it was shown that the rotational temperature of CH$_{3}$CCH gives a good estimate of the gas kinetic temperature at gas density $n\gtrsim 10^{3-4}$~cm$^{-3}$ (transitions $J=5-4$ and $J=6-5$ were considered). It is explained by the fact that, due to the low dipole moment ($\mu$ = 0.78~D), the CH$_{3}$CCH molecule is easily thermalized under such conditions. Gas density in our sources are above this threshold (Pazukhin et al., in preparation). Thus, the CH$_{3}$CCH lines in our data can be a good gas kinetic temperature indicator. Rotational (and, accordingly, kinetic) temperature is determined by the population diagrams method:}
\begin{equation}\label{eq:rotdi}
     \ln\left(\frac{3k \int T_{mb}d\upsilon }{8\pi^3 \nu \mu^2 g_Ig_KS}\right) = -\frac{E_u}{T_{kin}} + \ln\left(\frac{N_{tot}}{Q}\right),
\end{equation}
where $S$ is the line strength equal to $\frac{J^2 - K^2}{J}$, $\nu$ is the transition frequency, $E_u$ is the upper energy level {in temperature units}, $\mu$ is the dipole moment, $\int T_{mb}d\upsilon$ is the integrated line intensity, $N_{tot}$ is the total column density, Q is the partition function, $g_K$ is the $K$ degeneracy associated with the internal quantum number $K$ due to the projection of the total angular momentum onto a molecule axis,
$g_I$ is the statistical weight associated with the nuclear spin. {It is assumed here that the emission is optically thin and the background radiation can be neglected.}

\begin{figure}
\setcaptionmargin{5mm}
% \onelinecaptionsfalse % if the figure caption is multiline
\onelinecaptionstrue % if the figure caption is one-line
\captionstyle{normal}
    \begin{minipage}[h]{0.49\linewidth}
    \centering\includegraphics[width=1\linewidth]{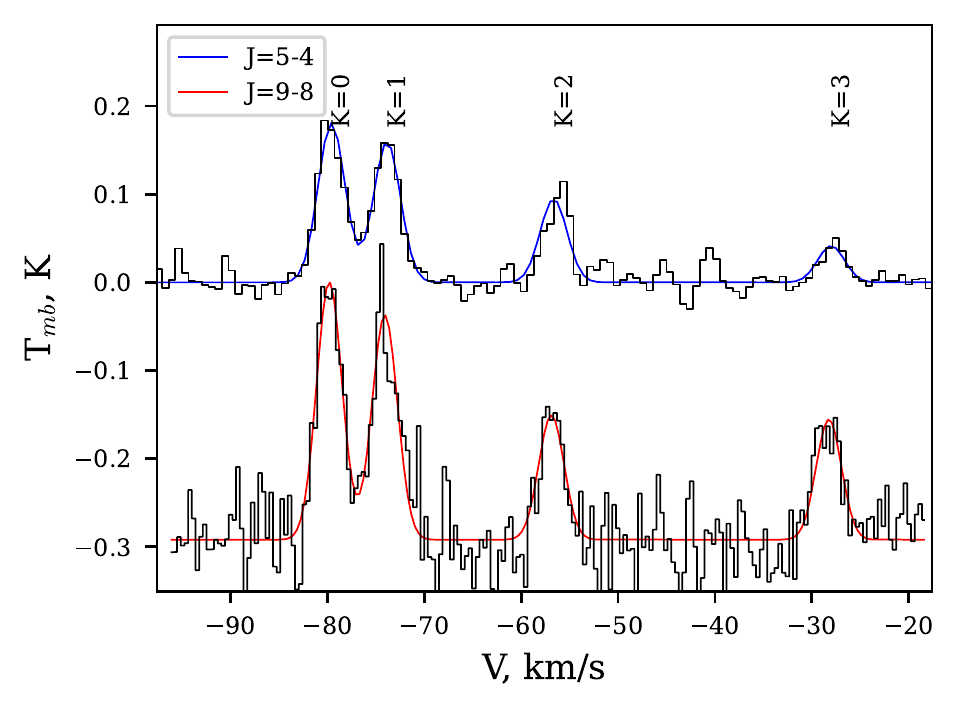}
    \end{minipage}
    \hfill
    \begin{minipage}[h]{0.49\linewidth}
    \centering\includegraphics[width=.85\linewidth]{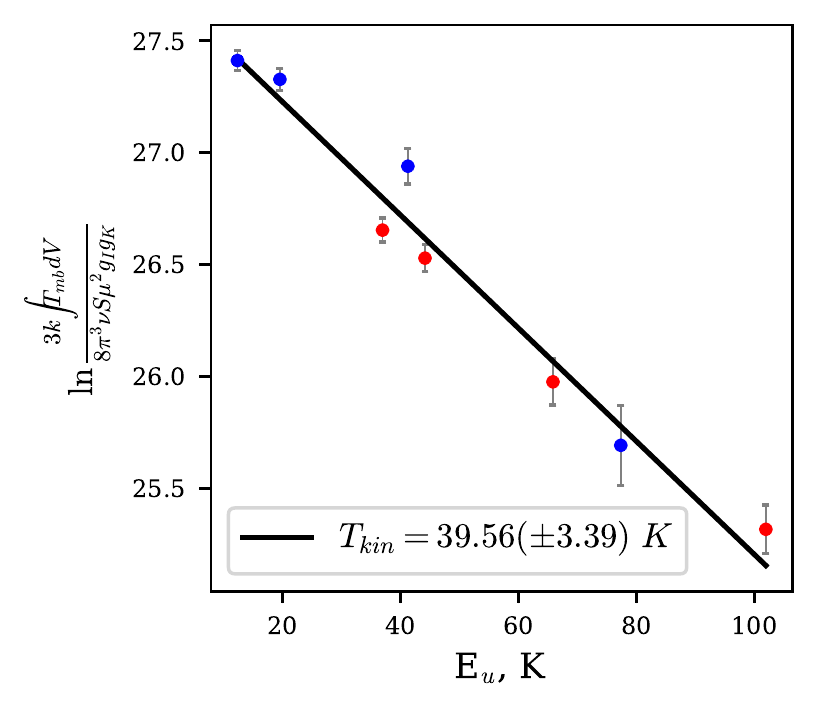}
    \end{minipage}
    \begin{minipage}[h]{0.49\linewidth}
    \centering\includegraphics[width=1\linewidth]{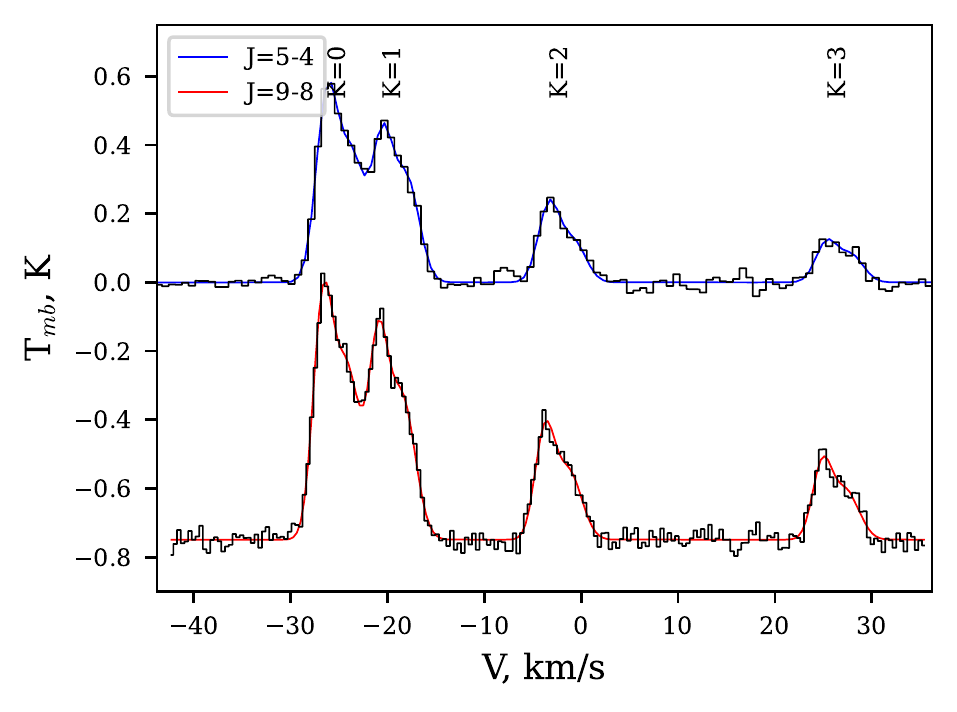}
    \end{minipage}
    \hfill
    \begin{minipage}[h]{0.49\linewidth}
    \centering \includegraphics[width=.85\linewidth]{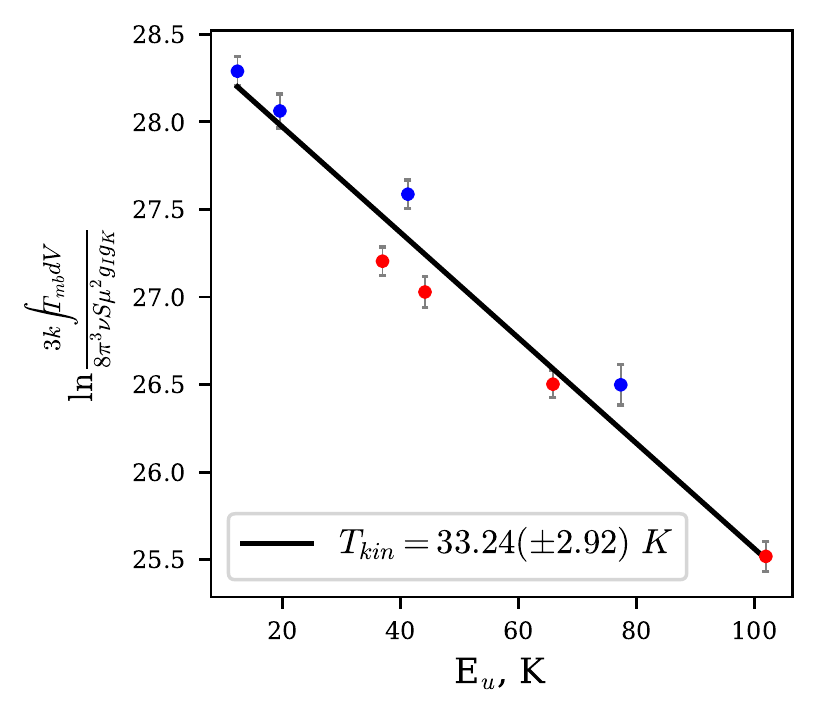}
    \end{minipage}
      \caption{Spectra (left) and population diagrams (right) for IRAS~23116+6111 (upper panels) and DR~21(OH) (lower panels). CH$_3$CCH $J = 9 - 8$ and $J = 5 - 4$ spectra are indicated in black; the red and blue lines are the Gaussian profile fitting. The lines in the population diagrams are plotted by the least squares method. In the lower left corner of the diagrams, the obtained value of the kinetic temperature is given. For DR~21(OH), the Gaussian profiles are plotted for the velocity components $\sim−4$  and $\sim0$~\kms, and the population diagram are plotted for the $\sim-4$~\kms\ component.}
      \label{fig:rd}
\end{figure}

The rotational diagrams were constructed using the $J=5-4$ and $J=9-8$ transitions of the CH$_{3}$CCH molecule. The spectra were fitted with Gaussian profiles, assuming that the widths of each component are equal, and the spacings between them are known. Then a graph was built, where the upper energy level $E_u$ was plotted along the abscissa axis, and the left part of the equation~(\ref{eq:rotdi}) was plotted along the ordinate axis. {Then,} $T_{kin}$ is proportional to the inverse of the slope. In Figure~\ref{fig:rd} in the direction IRAS~23116+6111 and DR~21(OH) the spectra of the CH$_{3}$CCH molecule and rotational diagrams are plotted. Figure~\ref{fig:Tk_compare}(left) shows a comparison of the kinetic temperature estimates for the $J=5-4$ and $J=9-8$ CH$_3$CCH transitions. In general {these estimates are close to each other}, therefore, the population diagram can be plotted using both transitions.

It should be noted that for the L~1287, estimates of the kinetic temperature were obtained only at two {positions} ($0^{\prime\prime}$, $0^{\prime\prime}$) and ($-14^{\prime\prime} $, $-14^{\prime\prime}$) and are equal to 21.5~$\pm$~1.9~K and 20.4~$\pm$~1.8~K, respectively. For S~187, S~231, the CH$_{3}$CCH lines turned out to be {too weak to estimate the kinetic temperature.}

\subsection{Kinetic temperature from NH$_3$ observations }

Transitions of the NH$_{3}$ molecule were observed in the sources S~187, DR~21(OH) with the {Effelsberg-100m} radio telescope. For S~231, we used the estimate of the kinetic temperature {with} ammonia from~\cite{Ryabukhina}.

The optical depth and rotational temperature were determined using the methods described in~\cite{Ho83}.
{The spectra were fitted with Gaussian profiles, in the transition (1,1) the widths of each component were {assumed to be} equal, and the spacings between them are known.}
Assuming that hyperfine components are under LTE conditions, the optical depth $\tau({1,1,m})$ can be determined from the ratio of the main and satellite line intensities:
\begin{equation}
    \frac{T^*_{A}({m})}{T^*_{A}({s}) }=\frac{1-\exp(-\tau({1,1,m}))} {1- \exp(-a\tau({1,1,m}))},
\label{eq:tau}
\end{equation}
where $T^*_{A}$ is the antenna temperature, $a$ is the ratio of the main and satellite line intensities, equal to $a=0.28$ for inner satellites and $a=0.22$ for outer satellites. The optical depth $\tau({1,1,m})$ was determined numerically from equation~(\ref{eq:tau}).
 
Thus, the rotational temperature can be obtained from the ratio of the main component intensities of (1,1) and (2,2) transitions using the equation:
\begin{eqnarray}
    T_{rot} = -41.5 \Bigg/ \ln \Big[ \frac{-0.282}{\tau(1,1,m)} \ln \Big( 1-\frac{T^*_{A}(2,2,m)}{T^*_{A}(1,1,m)} \{1-\exp(-\tau(1,1,m))\} \Big) \Big].
\label{eq:trot}
\end{eqnarray}

The kinetic temperature values were obtained using the equation from~\cite{Tafalla04}:
\begin{equation}
    T_{ kin} = \frac{ T_{ rot}}{ 1- \frac{T_{ rot}}{41.5} \ln \left[ 1+ 1.1 \exp \left( -\frac{16}{T_{ rot }} \right) \right] }.
\label{eq:tkin}
\end{equation}

\section{Discussion}

\begin{figure}
\setcaptionmargin{5mm}
% \onelinecaptionsfalse % if the figure caption is multiline
\onelinecaptionstrue % if the figure caption is one-line
\captionstyle{normal}
    \begin{minipage}[h]{0.49\linewidth}
        \center{\includegraphics[width=1\linewidth]{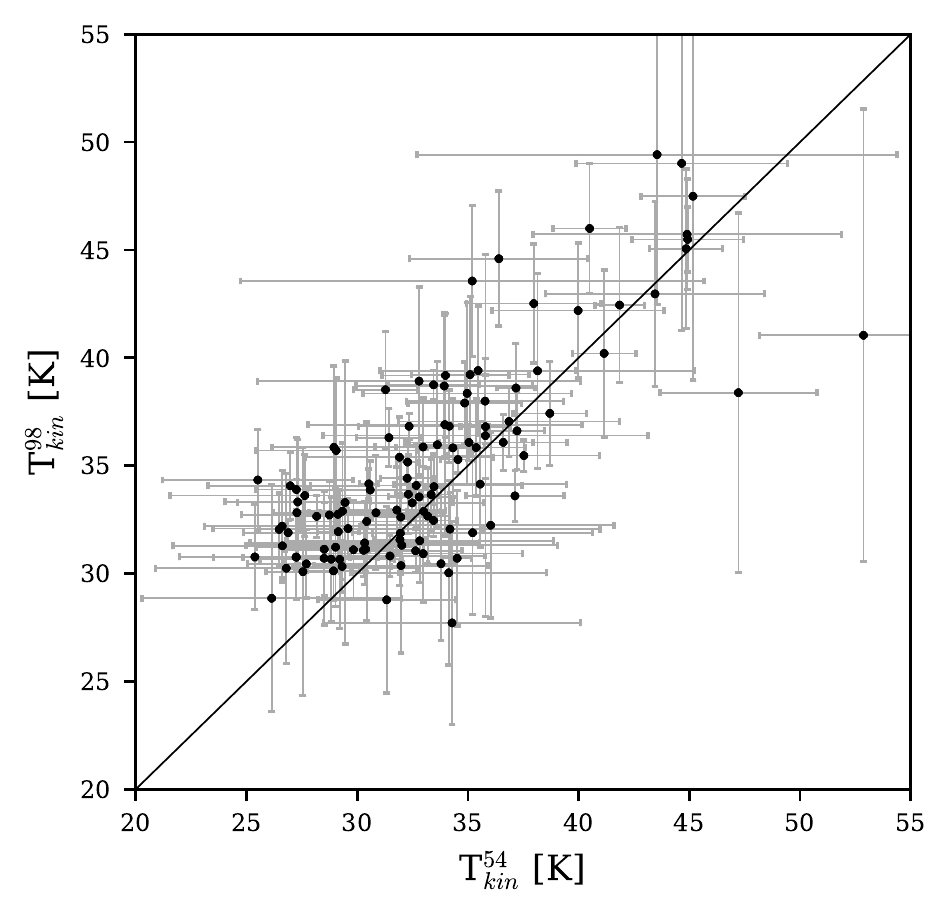}}
    \end{minipage}
    \begin{minipage}[h]{0.49\linewidth}
        \center{\includegraphics[width=1\linewidth]{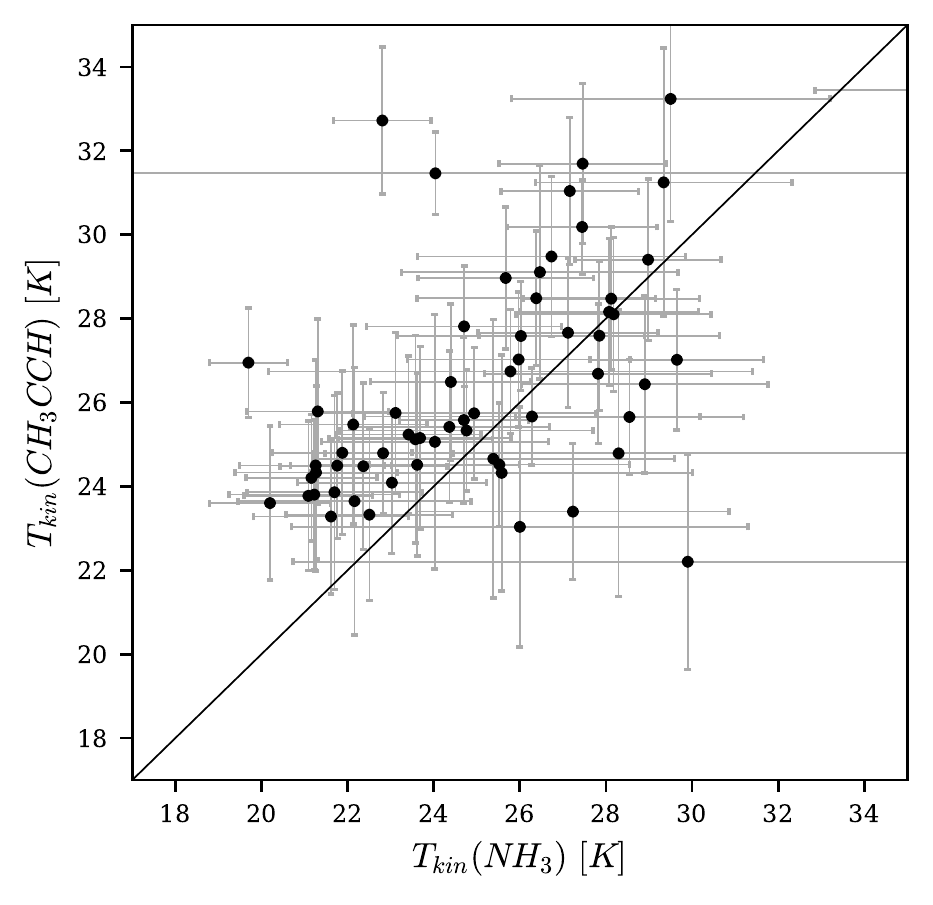}}
    \end{minipage}
    \caption{Comparison of estimates of the kinetic temperature from $J = 9 - 8$ and $J = 5 - 4$ CH$_3$CCH transitions (left) and for the source DR 21(OH) from ammonia and CH$_3$CCH transitions (right). Lines of the form $y = x$ are plotted diagonally.}
    \label{fig:Tk_compare}
\end{figure}

Figure~\ref{fig:Tk_compare}(right) shows a comparison of the kinetic temperature estimates from ammonia and CH$_3$CCH transitions for the DR~21(OH) source. {In general, there is a fairly good agreement between them, although the estimates} for the CH$_3$CCH transitions {result in} slightly higher values than the estimates for ammonia. This is probably due to the fact that methylacetylene is observed in a denser gas, where the temperature is higher.

\begin{figure}
\setcaptionmargin{5mm}
% \onelinecaptionsfalse % if the figure caption is multiline
\onelinecaptionstrue % if the figure caption is one-line
\captionstyle{normal}
    \center{\includegraphics{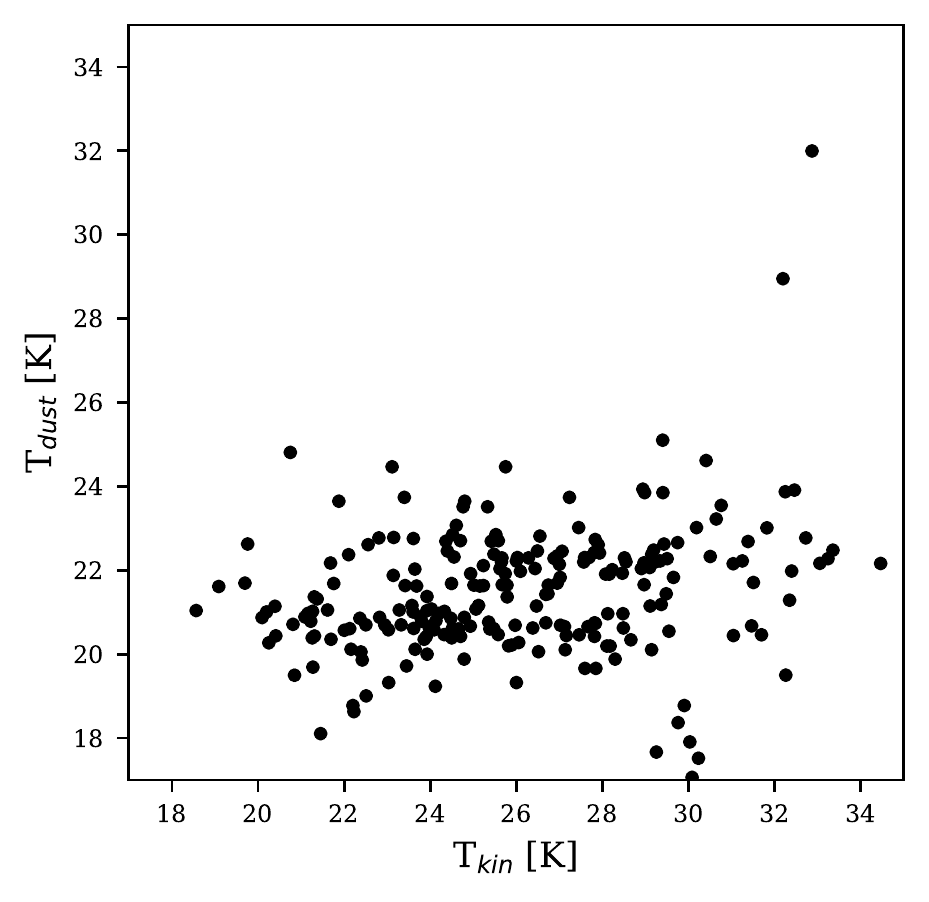}}
    \caption{Comparison of dust temperature and kinetic temperature.}
    \label{fig:Tk_Td}
\end{figure}

 In addition, for the sources L~1287, DR~21(OH) and NGC~7538, there are maps of dust temperature $T_{dust}$ and column density $N$(H$_2$) according to data from the {Herschel} telescope {taken} from the open database\footnote{\url{http://www.astro.cardiff.ac.uk/research/ViaLactea}}, which were obtained using the PPMAP~\cite{Marsh_1,Marsh_2} algorithm.
 We compared the dust temperature $T_{dust}$ estimates with the kinetic temperature estimates~(Fig.~\ref{fig:Tk_Td}). As a result, no significant correlation was found. $T_{dust}$ values are in the range $\sim18-25$~K, while $T_{kin}$ increase to 35~K. {It is possible that the lack of correlation is due to the insufficient density of our sources. Thus, in~\cite{Banerjee06} it is shown that the dust temperature approaches the gas temperature at a gas density $n\gtrsim 10^{7-8}$~cm$^{-3}$, which is much higher than the gas density estimates in our sources, which, according to our data, is $n\sim 10^{4-6}$~cm$^{-3}$ (Pazukhin et al., in preparation).}

\begin{figure}
\setcaptionmargin{5mm}
% \onelinecaptionsfalse % if the figure caption is multiline
\onelinecaptionstrue % if the figure caption is one-line
\captionstyle{normal}
    \begin{minipage}[h]{0.49\linewidth}
        \center{\includegraphics[width=1\linewidth]{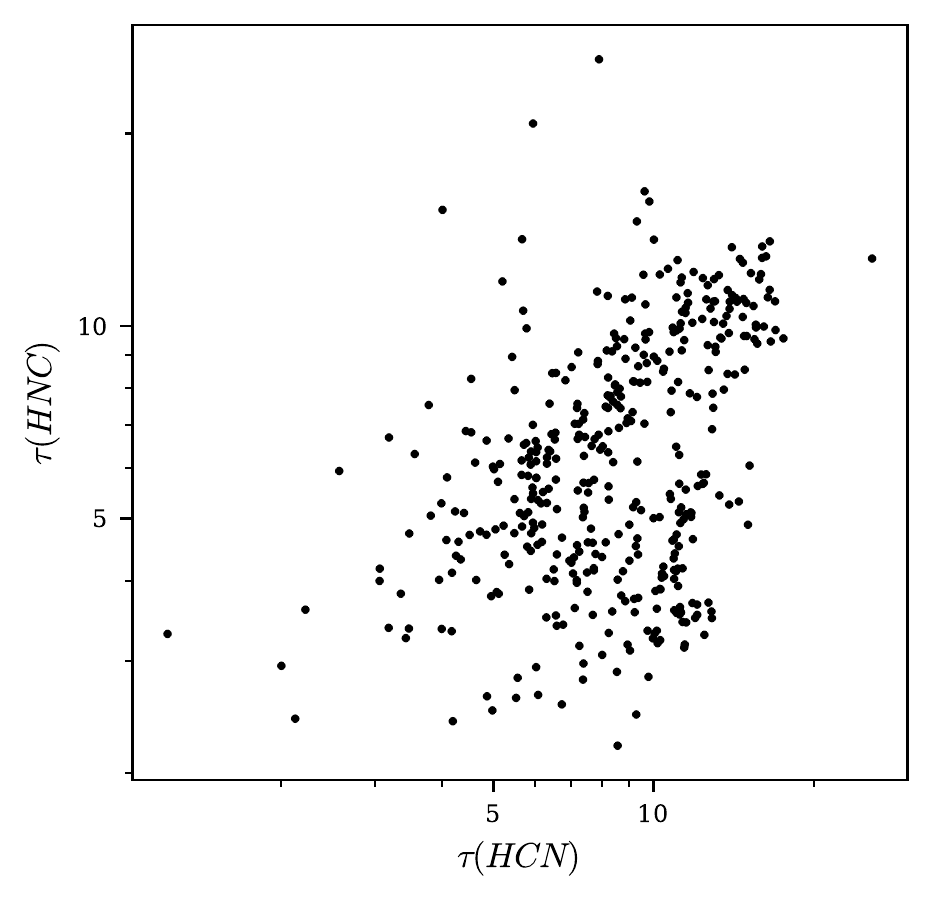}}
    \end{minipage}
    \begin{minipage}[h]{0.49\linewidth}
        \center{\includegraphics[width=1\linewidth]{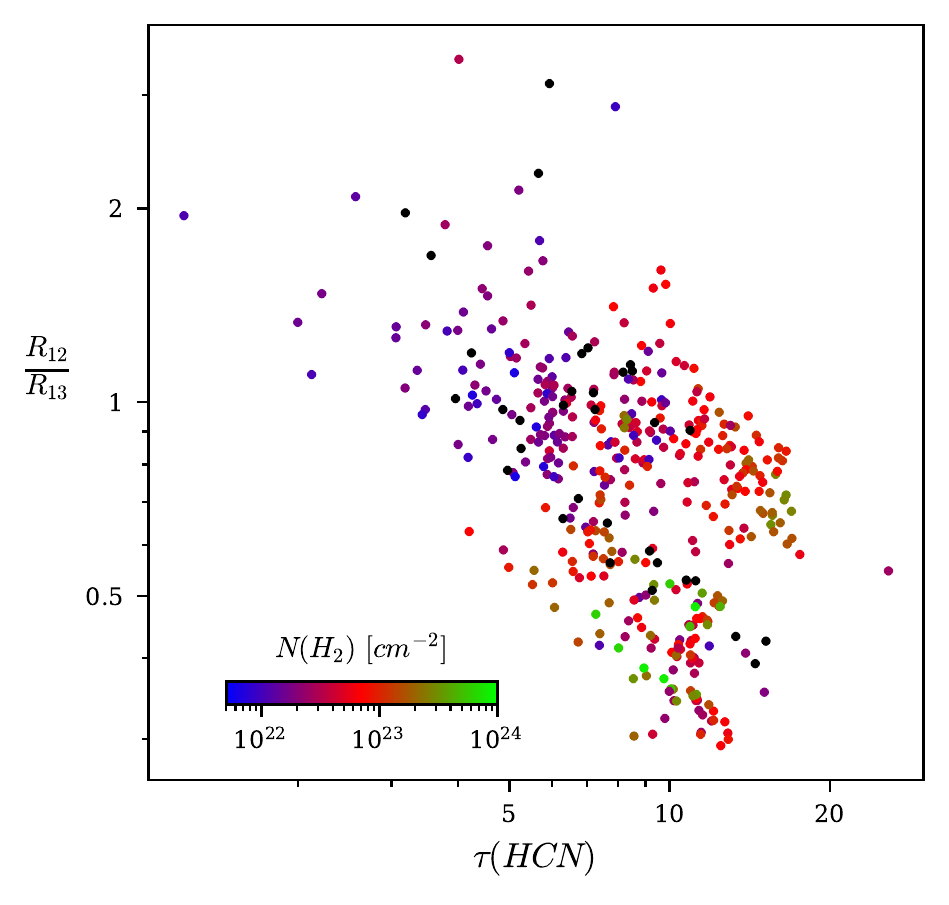}}
    \end{minipage}
    \caption{Dependence on the optical depth estimates $\tau \rm{(HCN)}$ for HCN and HNC molecules of the $\tau \rm{(HNC)}$ (left) and of the ratio $R_{12}/R_{13}$ (right). The color code indicates to the column density $N$(H$_2$) values.}
     \label{fig:tau}
\end{figure}

{We estimated the optical depths in the HCN and HNC lines. To do this, we used the intensity ratio of the isotopologues HCN/H$^{13}$CN and HNC/HN$^{13}$C in the equation (\ref{eq:tau}) and the value $a$ from the abundance ratio of carbon isotopes~\cite{Liu}
\begin{equation}
    \frac{^{12}\textrm{C}}{^{13}\textrm{C}}=4.7\times\textrm{R$_{GC}$}+25.05, \nonumber
\end{equation}
where R$_{GC}$ is the Galactocentric distance of the source.
Figure~\ref{fig:tau}(left) shows a comparison of the obtained optical depth values.
The optical depths in both lines are large. The optical depth in the HCN line is on average higher than in the HNC line and reaches $\sim20$. In this case, a rather large spread in the ratio of optical depths in these lines is observed. This makes it preferable to use the lines of their {rare} isotopologues H$^{13}$CN and HN$^{13}$C, in which the optical depth is obviously small.}
{In~\cite{Hacar20} it was came to the conclusion that the optical depth in the HCN and HNC lines does not significantly affect the relation between the intensity ratio of these lines and the gas temperature. Our data cast doubt on this.  Figure~\ref{fig:tau}(right) shows the dependence of the ratio $R_{12}/R_{13}$ [$R_{12}=I(\mathrm{HCN})/I(\mathrm{HNC })$ and $R_{13}=I(\mathrm{H^{13}CN})/I(\mathrm{HN^{13}C})$] on the optical depth $\tau \textrm{(HCN )}$, as well as the values of the column density $N(\textrm{H}_2)$. It can be seen that at large optical depths, which are typical for HCN lines, the ratio $R_{12}/R_{13}$ is much lower than unity. As the optical depth decreases, this ratio, as expected, tends to 1. The hydrogen column density, at which the optical depth in the lines becomes small, is $N(\mathrm{H}_2) \sim 10^{22}$~cm$^{-2 }$.}

\begin{figure}
\setcaptionmargin{5mm}
% \onelinecaptionsfalse % if the figure caption is multiline
\onelinecaptionstrue % if the figure caption is one-line
\captionstyle{normal}
    \center{\includegraphics{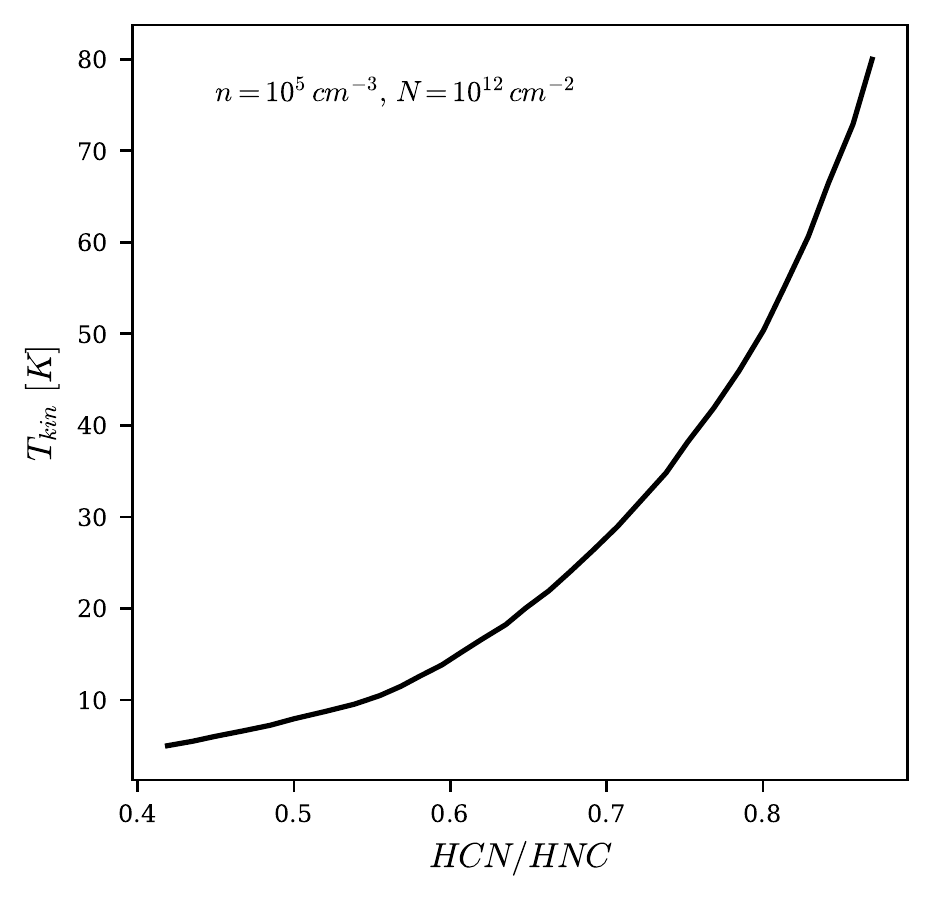}}
    \caption{Dependence of the kinetic temperature $T_{kin}$ on the ratio $R_{12}$ plotted using the RADEX program for $n=10^{5}$~cm$^{-3}$ and $N=10^{ 12}$~cm$^{-2}$.}
    \label{fig:Tex}
\end{figure}

 {Variations of the $R_{12}$ ratio can be caused by different excitation temperatures $T_{ex}$ of HCN and HNC. However, on the dependence between $T_{kin}$ and $R_{12}$ plotted using the RADEX~\cite{Radex} program under the conditions $n=10^{5}$~cm$^{-3}$ and $N=10^{12}$~cm$^{-2}$ it can be seen that the ratio {changes only slightly} with increasing temperature and amounts to $\alt 1 $~(Fig.~\ref{fig:Tex}).}

\begin{figure}
\setcaptionmargin{5mm}
% \onelinecaptionsfalse % if the figure caption is multiline
\onelinecaptionstrue % if the figure caption is one-line
\captionstyle{normal}
    \begin{minipage}[h]{0.49\linewidth}
        \center{ \includegraphics[width=1\linewidth]{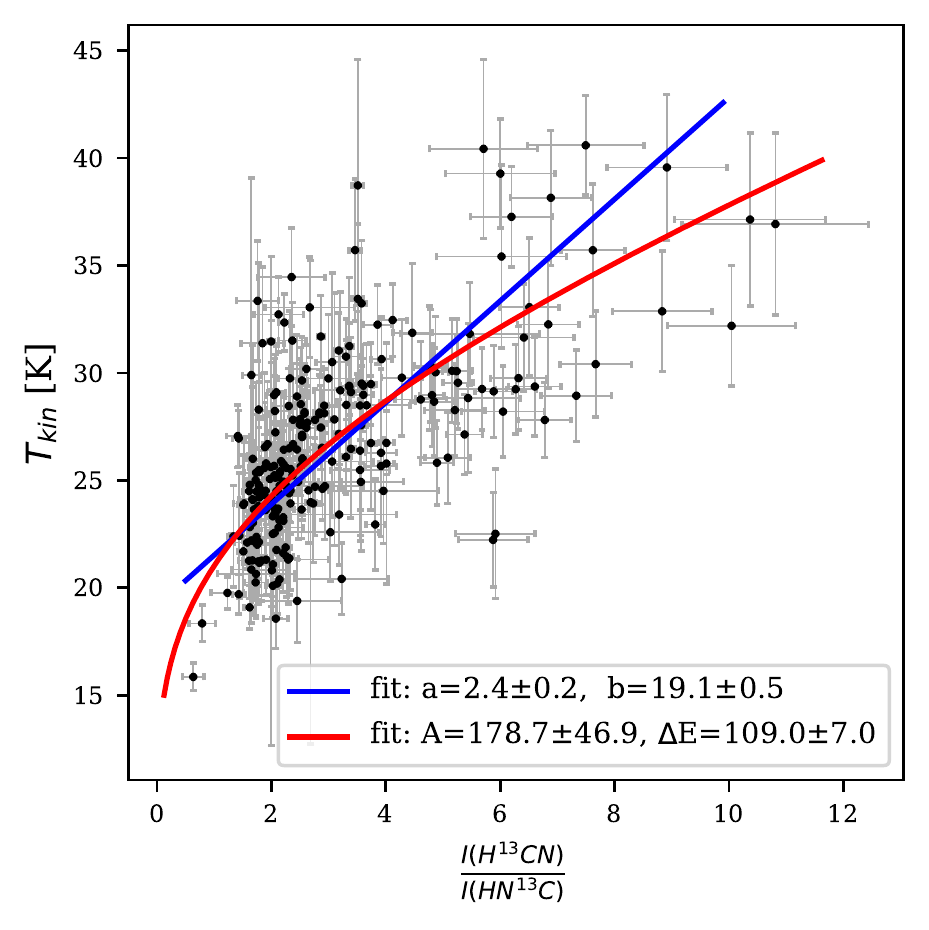}}
    \end{minipage}
    \begin{minipage}[h]{0.49\linewidth}
        \center{ \includegraphics[width=1\linewidth]{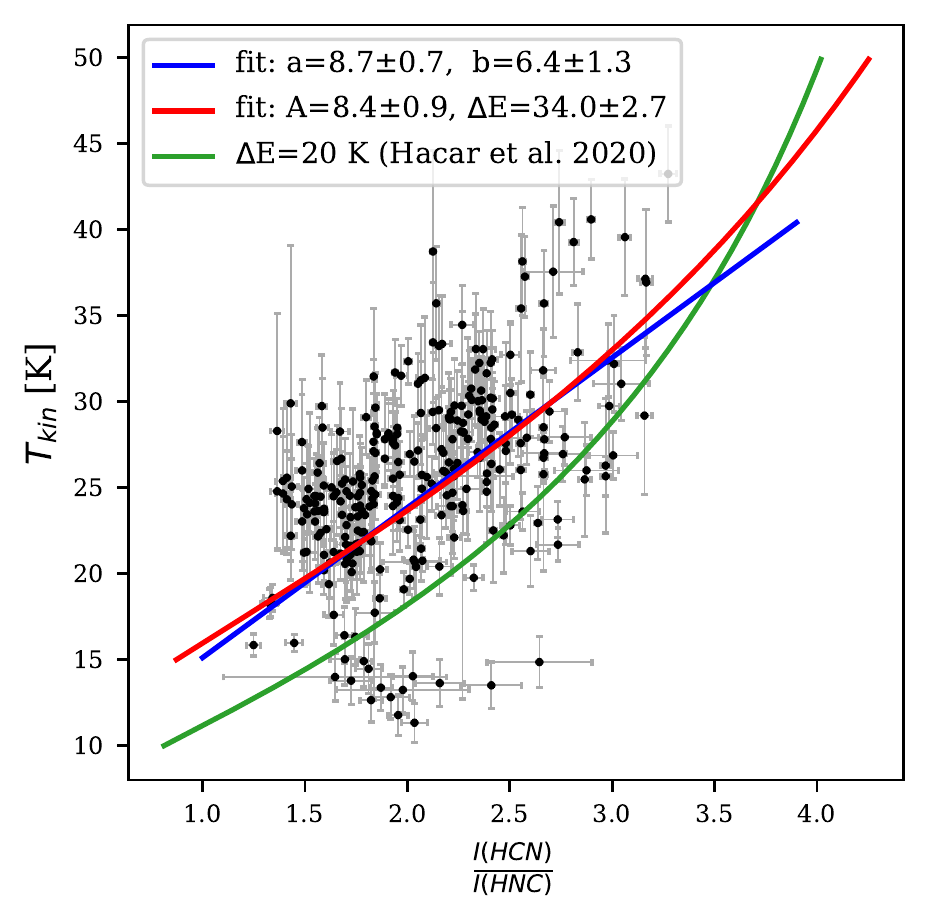}}
    \end{minipage}
    \caption{Dependence of the kinetic temperature on the integrated intensity ratio of the molecules H$^{13}$CN and HN$^{13}$C (left) and HCN and HNC (right). The fitting results are represented by the {blue straight  line $ax + b$} and the red curve describes a function $A \exp\left( {\frac{-\Delta E}{T_{kin}}} \right)$. The green curve corresponds to $\Delta E = 20$~K from~\cite{Hacar20}. The parameters of {the fits} ($a$, $b$, $A$, $\Delta E$) are shown in each of the figures.}
    \label{fig:rat_Tk}
\end{figure}

As a result of the analysis of our data, the dependence of the ratios $R_{12}$ and $R_{13}$ on the gas kinetic temperature was plotted (Fig.~\ref{fig:rat_Tk}). The value of $R_{13}$ increases from 1 to 10, and the intensity ratio of the main isotopologues increase from 1 to 4 in the temperature range $\sim15-45$~K.

Thus, as a result of the linear least squares fit, the following dependencies were obtained:
\begin{eqnarray}
    T_{kin} =
    \begin{cases}
    2.4\times R_{13} + 19.1,\\
    8.7\times R_{12} + 6.4.
    \end{cases}
    \label{eq:rat_Tk}
\end{eqnarray}

The line obtained for HCN and HNC agrees with the line $T_{kin}=10\frac{\rm{I(HCN)}}{\rm{I(HNC)}}$ obtained in~\cite{Hacar20}, which is valid for intensity ratios $\leq4$ and up to temperatures $T_{kin}\sim40$~K.

In addition, Figure~\ref{fig:rat_Tk} shows an approximation by a function of the form $A \exp\left( {\frac{-\Delta E}{T_{kin}}} \right)$, which is chosen based on the population ratio expressed in terms of the Boltzmann distribution. As a result, {the following dependencies were}:
\begin{subequations}
\begin{eqnarray}
    R_{13} = 179\times \exp \left( {\frac{-109}{T_{kin}}} \right), \label{eq1:rat_dE} \\
    R_{12} = 8.4\times \exp \left( {\frac{-34}{T_{kin}}} \right).
    \label{eq2:rat_dE}
\end{eqnarray}
\end{subequations}
The energy barrier for the ratio $R_{13}$ is $\Delta E \sim109$~K, and for the main isotopologues is $\Delta E \sim34$~K.
The results from other publications are somewhat different, the energy barrier at low temperatures is $\Delta E \sim20$~K~\cite{Hacar20}, with a further increase with temperature $\Delta E \sim200$~K~\cite{Hirota,Graninger}.

In general, the data for HCN and HNC agree with the results from~\cite{Hacar20}. However, the results for $R_{13}$ are noticeably different. The main reason for the discrepancy between the results is probably the large optical depth of the HCN and HNC lines, as well as the presence of anomalies in the hyperfine structure of the HCN molecule.
 
{The use of the H$^{13}$CN and HN$^{13}$C lines for temperature estimation was also recently proposed and demonstrated in~\cite{Beuther22}. However, in this paper, to estimate the temperature, the correlation dependence of $R_{12}$ on temperature found in~\cite{Hacar20} is used. As shown above, the dependence of $R_{13}$ on temperature differs from it.}

\begin{figure}
\setcaptionmargin{5mm}
% \onelinecaptionsfalse % if the figure caption is multiline
\onelinecaptionstrue % if the figure caption is one-line
\captionstyle{normal}
    \begin{minipage}[h]{1\linewidth}
        \center{\includegraphics[width=1\linewidth]{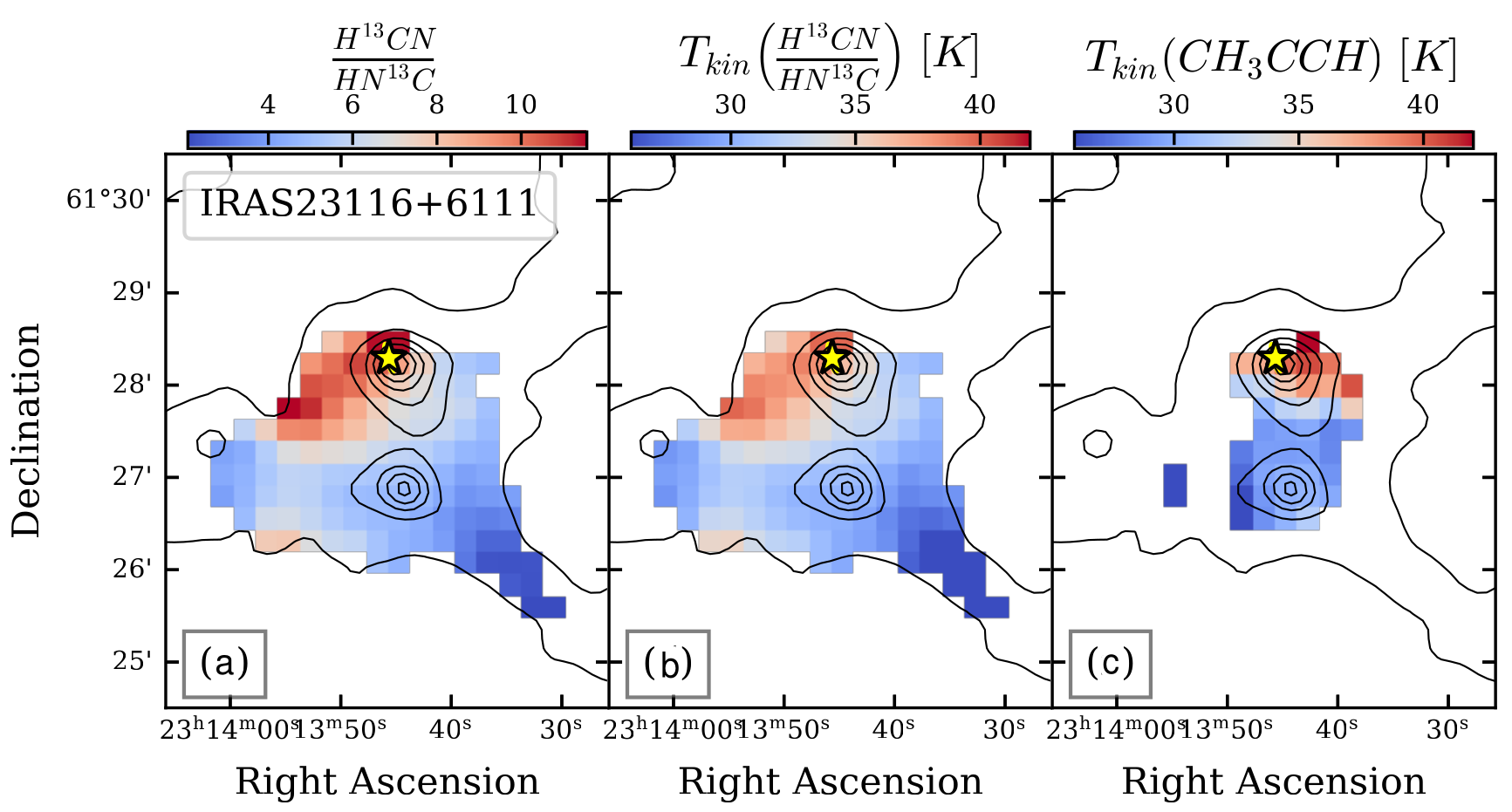} }
    \end{minipage}
    \begin{minipage}[h]{1\linewidth}
        \center{\includegraphics[width=1\linewidth]{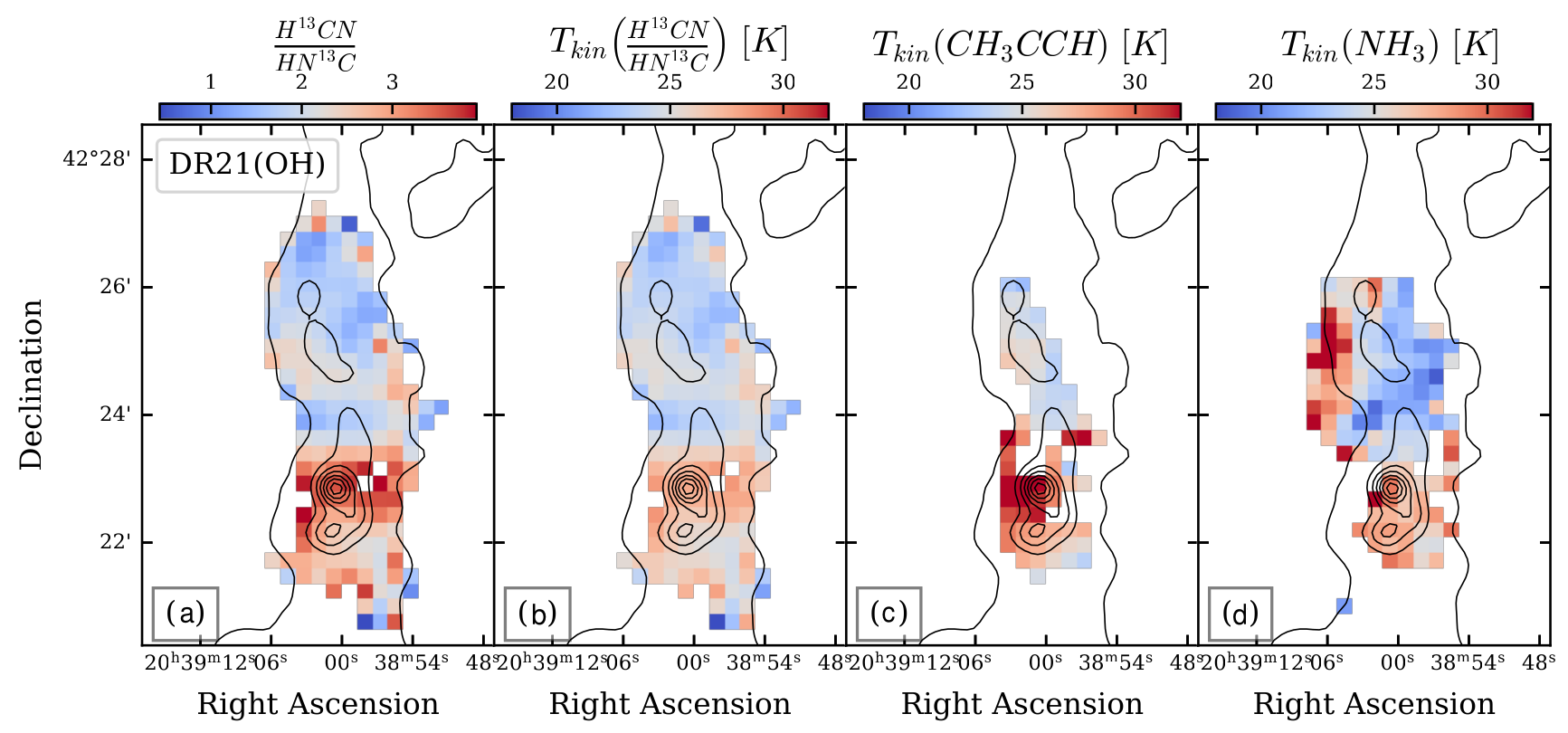} }
    \end{minipage}
      \caption{Maps for NGC~7538 (upper panels) and DR~21(OH) (lower panels). The integrated intensity ratio of the molecules H$^{13}$CN and HN$^{13}$C~(a); $T_{kin}$(H$^{13}$CN/HN$^{13}$C)~(b), the kinetic temperature derived using CH$_3$CCH transitions~(c) and the kinetic temperature derived using NH$_3$ transitions~(d). The SCUBA~850~$\mu m$ continuum data~\cite{scuba} are indicated by contours. The star-shaped marker indicates the IRAS~23116+6111 position.}
\label{fig:maps}
\end{figure}

We suppose that it is preferable to use the equation~(\ref{eq1:rat_dE}) for the ratio $R_{13}$ as a temperature indicator. Temperature estimates for  NGC~7538 and DR~21(OH) are shown in~Figure~\ref{fig:maps}. The plotted maps demonstrate good agreement with the estimates obtained from the CH$_3$CCH and NH$_3$ lines. The temperature gradient is visible, {the peaks spatially coincide with the emission of the continuum} and the emission of the IR source. In addition, maps {are more extended than} than plotted temperature maps obtained from lines of CH$_3$CCH and NH$_3$.

{It is worth noting that temperature maps can be further expanded by combining the observational data from isotopologues H$^{13}$CN and HN$^{13}$C with observations from the main isotopologues, for example, as suggested in~\cite{Beuther22}. In this paper, in those source regions where the H$^{13}$CN and HN$^{13}$C lines become too weak, the intensity ratio of the main isotopologues $R_{12}$ is used.}

\section{Conclusion}

Based on observations of five massive star-forming regions obtained with the {IRAM-30m} and {Effelsberg-100m} radio telescopes, as well as using estimates of the dust temperature $T_{dust}$ from the {Herschel} telescope data, we obtained following results:
\begin{enumerate}
    \item A correlation between the integrated intensity ratios of the $J=1-0$ transitions of H$^{13}$CN and HN$^{13}$C and the kinetic temperature has been found.
    The intensity ratio increases from 1 to 10 in the temperature range $\sim15-45$~K. Since these lines can be detected in observations of most sources, the results obtained allow us to propose using the intensity ratio H$^{13}$CN/HN$^{13}$C as a possible temperature indicator of interstellar clouds.
    \item For the low-temperature reaction HNC+O$\to$CO+NH, the energy barrier obtained from the ratio H$^{13}$CN/HN$^{13}$C was $\Delta E \sim109$~K, and from the ratio of the main isotopologues $\Delta E \sim34$~K.
    The main reason for the discrepancy between the results is probably the large optical depth of the HCN and HNC lines, as well as the presence of anomalies in the hyperfine structure of the HCN molecule.
    \item We compared the obtained estimates of the kinetic temperature with the estimates of the dust temperature $T_{dust}$. As a result, no significant correlation was found. $T_{dust}$ values are in the range $\sim18-25$~K, while $T_{kin}$ grows up to 35~K.
    It is possible that the lack of correlation is due to the insufficient density of the observed sources.
\end{enumerate}

\section*{Funding}

The research was supported by the Russian Science Foundation (grant no. 22-22-00809).

\begin{acknowledgments}
The research is based on observations made by the 041-19 project with the 30-m telescope, as well as observations with the 100-m MPIfR telescope (Max-Planck-Institut für Radioastronomie) in Effelsberg. IRAM is supported by INSU/CNRS (France), MPG (Germany) and IGN (Spain). We acknowledge the staff of both observatories for their support in the observations. {The authors are grateful to the anonymous reviewer for useful comments that improved the quality of the paper.}

\end{acknowledgments}

%
% Bibliography
%
\bibliographystyle{maik}
\bibliography{pazukhin}

\end{document}